\documentclass[aps,prx,twocolumn,showpacs,superscriptaddress,longbibliography]{revtex4-1}
\usepackage[bookmarks=false,linkcolor=blue,urlcolor=blue,colorlinks,citecolor=blue]{hyperref}

\usepackage{graphicx}
\usepackage{color}
\usepackage{amsmath}
 \usepackage{setspace} 
\usepackage[normalem]{ulem}

\include{bibliography}

\begin{document}

\title{Non-quasiparticle transport and resistivity saturation: A view from the large-N limit}

\author{Yochai Werman}
\affiliation{Department of Condensed Matter Physics, The Weizmann Institute of Science, Rehovot, 76100, Israel}
\author{Steven A. Kivelson}
\affiliation{Department of Physics, Stanford University, Stanford, CA 94305, USA}
\author{Erez Berg}
\affiliation{Department of Condensed Matter Physics, The Weizmann Institute of Science, Rehovot, 76100, Israel}

\date{\today}

\begin{abstract}

The electron dynamics in metals are usually 
well described  by the semiclassical approximation 
for long-lived quasiparticles. 
However, in some metals, the scattering rate of the electrons at elevated temperatures becomes comparable to the Fermi energy; then, this approximation breaks down, and the full quantum-mechanical nature of the electrons must be considered. In this work, we study a solvable, large-$N$ electron-phonon model, which at high temperatures enters the non-quasiparticle regime. In this regime, the model exhibits ``resistivity saturation'' to a temperature-independent value of the order of the quantum of resistivity - the first analytically tractable model to do so. 
The saturation is not due to a fundamental limit on the electron lifetime, but rather to the appearance of a 
second conductivity channel. This 
is suggestive of the phenomenological ``parallel resistor formula'', known to describe the resistivity of a variety of saturating metals.

\end{abstract}

\maketitle

{\emph{Introduction.--}  The  tendency for the resistivity of metals to increase with  temperature, $T$, 
is generally understood on the basis of Boltzmann transport theory. In turn, for the requisite distribution function to be consistent with quantum mechanics, it must be possible to construct electron wave-packets with well defined velocities and positions.  Consequently, at best, Boltzmann  theory is applicable only so long as the mean-free-path, $\ell$, is long compared to the Fermi wavelength, i.e.  $\ell \gg  2\pi/k_F$.  There are other conditions for the validity  of Boltzmann theory, such as the Mott-Ioffe-Regel (MIR) condition $\ell \gg a$,
 which allows one to ignore interband scattering.  
While there is no upper bound on the magnitude of a metallic resistivity, any time $\rho \gtrsim \rho_{\mathrm{B}}/N$ (where $N$ is the number of bands), it cannot be interpreted in terms of  freely propagating quasiparticles which are occasionally scattered.  Here $\rho_{\mathrm{B}}$ signifies a 
characteristic resistivity derived from Boltzmann theory extrapolated to the limit $\ell = 2\pi/k_F$, i.e. $\rho_{\mathrm{B}}\equiv \hbar/e^2$ in $d=2$ and  $\rho_{\mathrm{B}}=(4/3) [h/e^2 k_F]$ in $d=3$, while the 
MIR limit ~\cite{Ioffe} corresponds to $\rho_{\mathrm{MIR}} = h a^{d-2}/e^2$.

In practice, many simple metals  melt before $\rho$ gets to be as large as $\rho_{\mathrm{B}}$.  Of those that reach this value, there are apparently two distinct classes:  a) Those that exhibit ``resistivity saturation,'' i.e. the resistivity becomes decreasingly $T$ dependent as $T$ gets large, with a value that appears to approach a finite asymptotic limit at large $T$.  b)  Those ``bad metals''~\cite{Kivelson} for which $\rho_{\mathrm{B}}$ does not appear to be a relevant scale at all, in which the resistivity is still a strongly increasing function of $T$ even when $\rho > \rho_{\mathrm{B}}$.}
%
Understanding bad metallic behavior, and its 
complement, resistivity saturation, remains one of the major open problems in the theory of metals~\cite{Kivelson,Hussey,Gunnarsson}. 
 Transport regimes beyond the quasi-particle paradigm have attracted much interest in recent years~\cite{Mukerjee2006, Varma2009, 
 Lindner2010, Hartnoll2011, Wolfle2011, Kotliar, deng2013bad, Syzranov2012, Mahajan2013, Hartnoll2014, pakhira2015absence,Limtagool2015, Hartnoll}. 


Since its discovery in the 1970s~\cite{FiskLawson,FiskWebb, savvides1982electrical}, several theories have been proposed to explain resistivity saturation~\cite{chakraborty1979boltzmann, Allen1981, Auerbach1984, Calandra, WegerMott, Weger, Laughlin, Cote, Christoph, Millis, Allen2002}; however, to this day, no consensus has emerged. In particular, several key theoretical issues have not been resolved; for example, in cases where $2\pi/k_F$ and $a$ are parametrically different from each other (as in a weakly doped semiconductor), it is not clear whether the saturation value of the resistivity corresponds to 
$\ell\approx 2\pi/k_F$, 
$\ell\approx a$, or neither. 
Empirically, the resistivity of saturating metals is 
often well-described by the ``parallel resistor'' formula~\cite{Wiesmann},
\begin{eqnarray}
\label{eq:parallel}
\rho(T)^{-1}=\rho_{\mathrm{ideal}}(T)^{-1}+\rho_{\mathrm{sat}}^{-1},
\end{eqnarray}
with $\rho_{\mathrm{ideal}}(T)=\rho_0+\gamma T$ representing the semiclassical contribution of disorder and phonon scattering (where $\rho_0$ and $\gamma$ are constants), and $\rho_{\mathrm{sat}}$ the saturation resistivity. This formula suggests the existence of a parallel conduction channel which is not affected by phonon scattering. Moreover, it is typically 
the case that $\rho_{\mathrm{sat}}\sim \rho_{\mathrm{B}}\sim\rho_{\mathrm{MIR}}$. 


In this paper we present a tractable microscopic electron-phonon model with a resistivity that saturates at a value 
$\rho_{\mathrm{sat}}$ that is {\em independent of the strength of the electron-phonon coupling}, but that does depend on the electron density and the band structure;  in that sense, while numerically it is not all that different from either $\rho_{\mathrm{B}}$ or $\rho_{\mathrm{MIR}}$, conceptually it does not quite correspond to either.
In addition, two distinct conductivity channels appear: one which continuously decreases with increasing temperature, and another which saturates at high temperatures. This is reminiscent of the parallel resistor formula, Eq.~(\ref{eq:parallel}). In 
this model, the saturation of resistivity is not due to a bound on the quasiparticle 
lifetime or 
 its mean free path, but on the existence of a $T$-independent phonon-assisted conduction channel.  

\begin{figure}[h!]
\centering
\includegraphics[width=0.45\textwidth]{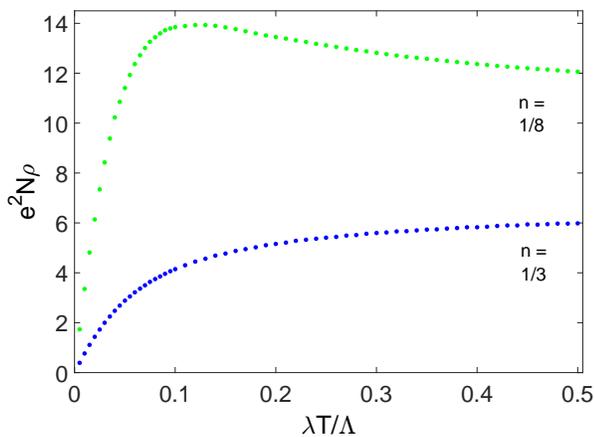}
  \caption{(Color online.) Resistivity per flavor, { in units of $1/e^2N$}, as a function of $\lambda T/\Lambda$ {in the $N\to \infty$ limit}, where $\Lambda$ is the bandwidth, for a two-dimensional square lattice. 
 The blue and green curves are, respectively, for a density per site in each flavor, $n=1/3$ and $n=1/8$. 
 at the lower density, the resistivity exceeds the saturating value and then approaches it with increasing temperature from above.}
\label{fig:ressat}
  \centering
    \end{figure}

Our model consists of $N$ identical electronic bands coupled to $N^2$ optical (Einstein) phonon modes. 
As in Ref.\cite{Werman}, we consider the problem in the limit that the dimensionless electron-phonon coupling (defined in Eq. \ref{eq:c} below) is large, $\lambda \gg 1$.  
 This is a necessary condition to insure that Boltzman transport theory breaks down at a temperature, $T_B \sim E_F/\lambda$, that is small compared to the Fermi energy $E_F$. 
 We shall see that at low electron density, where the bandwidth $\Lambda \gg E_F$, it is possible to look separately at the crossover that occurs at $T_B$ and at $T_{MIR} \sim \Lambda/\lambda$, while still maintaining $T \ll E_F$.  
There are generically many unwanted complications, including possible lattice instabilities, associated with large $\lambda$;  however, the combination of the large $N$ limit taken here, and the fact that we are studying phenomena at relatively high temperatures makes them irrelevant in the present study.~\footnote{In the electron-phonon coupling, we keep only linear order in the phonon displacement, and neglect higher-order terms. This is justified 
  in the large-$N$ limit, 
 since the typical magnitude of the coupling term to any single phonon mode is $\alpha \sqrt{T/(K\, N)}$, which is smaller than other electronic scales (e.g., $E_F$)} 
Representative results for the resistivity as a function of temperature are shown in Fig.~\ref{fig:ressat}. 

\begin{figure}	
\centering
\includegraphics[width=0.4\textwidth]{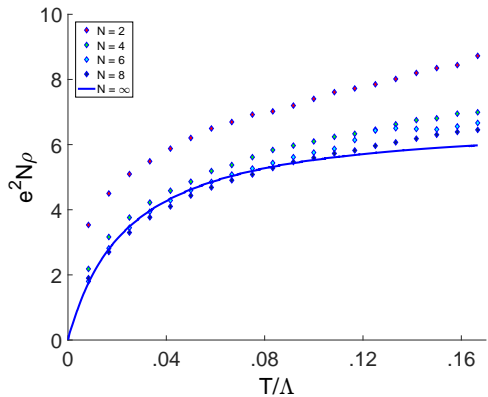}
  \caption{(Color online) Numerical results for the resistivity as a funciton of temperature. These results were obtained using a Monte Carlo simulation, treating the phonons classically and the electrons quantum mechanically.}
\label{fig:num}
  \centering 
\end{figure}

In order to verify that the behavior we find is not an artifact of the large $N$ limit, we have performed numerical Monte-Carlo simulations of the model at finite values of $N$. The results (see Figure~\ref{fig:num}) 
confirm that the qualitative behavior of the $N\rightarrow \infty$ solution 
are already apparent for $N$ as small as four.

\emph{Model.--}
Our system is composed of $N\gg1$ electron bands, which interact with $N^2$ optical, dispersionless phonon modes, in $d$ spatial dimensions. The phonons couple to the electron kinetic energy. In this type of large-$N$ expansion, inspired by the work of Fitzpatrick et al.~\cite{Raghu}, the phonon modes act as a momentum and energy bath for the electrons; thus, it is particularly suitable for studying the effects of the phonons on the electrons, while neglecting the back action of the electrons on the phonons. 
(This is probably a reasonable assumption in the relevant temperature range even for ``realistic'' small values of $N$.)

The action is given by
\begin{equation}\label{eq:model}
S = S_{\mathrm{el}} + S_{\mathrm{ph}} + S_{\mathrm{int}},
\end{equation}
where
\begin{equation}
S_{\mathrm{el}}=\sum_{a=1}^N \sum_{\nu_n}\int\frac{ d^dk}{(2\pi)^d} c^\dagger_a(\mathbf{k},\nu_n)\left[i \nu_n-\xi_{\mathbf{k}}\right] c_a(\mathbf{k},\nu_n)
\end{equation}
is the electronic part of the action,
\begin{equation}
S_{\mathrm{ph}} = \sum_{a,b=1}^N \sum_{\omega_n,r}\int \frac{d^dq}{(2\pi)^d}\frac{1}{2}\left[M\omega_0^2+M\omega_n^2\right]|X^r_{ab}(\mathbf{q},\omega_n)|^2
\end{equation}
is the phononic part, and 
\begin{eqnarray} \label{eq:Lint}
S_{\mathrm{int}} &=& \frac{\alpha}{\sqrt{\beta N}}\sum_{a,b=1}^N\sum_{\nu_n,\nu_m, r}\int \frac{d^dkd^dk'}{(2\pi)^{2d}}g_r(\mathbf{k},\mathbf{k}') \\
&\times& X^r_{ab}(\mathbf{k}-\mathbf{k'},\nu_n-\nu_m)\left[c^\dagger_a(\mathbf{k},\nu_n)c_b(\mathbf{k'},\nu_m)+a\leftrightarrow b\right]\nonumber
\end{eqnarray}
is the electron-phonon interaction term.
Here, $c^\dagger_a(\mathbf{k},\nu_n)$ creates an electron of wavevector $\mathbf{k}$, Matsubara frequency $\nu_n$, and flavor $1 \le a \le N$; the electronic dispersion is $\xi_{\mathbf{k}}=\epsilon(\mathbf{k})-\mu(T)$, with $\epsilon(\mathbf{k})\in [-\Lambda/2,\Lambda/2]$ (where $\Lambda$ is the bandwidth). $\mu(T)$ is the chemical potential at temperature $T$, and $\beta=1/T$. $X^r_{ab}(\mathbf{q},\omega_n)$ is the Fourier transform of the phonon displacement operator of flavor $a,b$ and mode $r$; $M$ is the ionic mass, and $\omega_0$ is the phonon frequency. $\alpha$ is the electron-phonon coupling strength. The dimensionless form factor $g_r(\mathbf{k},\mathbf{k}')$ satisfies $g_r(\mathbf{k}',\mathbf{k})=g_{r}(\mathbf{k},\mathbf{k}')^*$. 
Throughout the paper we set $k_B$, $\hbar$ and the lattice spacing $a$ to $1$. 

For concreteness, we use a $d=2$ tight binding model on a square lattice; we expect the results to be {qualitatively} insensitive to this particular choice. { We consider a case where the phonons couple to the electron bond density (as in the Su-Schrieffer-Heeger model~\cite{SSH}). There is one phonon mode centered on every bond; we label the phonon modes by the direction of the bond, $r = x,y$. The electron-phonon coupling term has the form} 
\begin{eqnarray}
\frac{\alpha}{\sqrt{4\beta N}}\sum_{a,b,j,r}X^r_{ab,j}\left(c^\dagger_{a,j} c^{\vphantom{\dagger}}_{b,j+r}+h.c.+a\leftrightarrow b\right),
\end{eqnarray}
where $j$ labels lattice sites. This term describes coupling of the phonon modes to the electron bond density. The corresponding electron-phonon form factor in Eq.~(\ref{eq:Lint}) is $g_r(\mathbf{k},\mathbf{k}') = e^{ik_ra}+e^{-ik'_ra}$. This is in contrast to the models studied by Millis et al.~\cite{Millis} and by us~\cite{Werman}, where the phonons couple to the electron site density. {The electronic dispersion is given by $\epsilon(\mathbf{k}) = -2t\sum_r \cos(k_r)$.} 

{As in Ref.~\cite{Werman}, 
we focus on the range of temperatures $\omega_0 \ll T \ll E_F$, where $\omega_0$ is the mean optical phonon frequency and $E_F$ is the Fermi energy.  The first inequality implies that the phonon variables can be treated as classical (our results are accurate to leading order in $\omega_0/T$), and the second that the electron fluid is still highly quantum mechanical.}


We define the dimensionless electron-phonon coupling constant as
\begin{eqnarray}
\label{eq:c}
\lambda=\frac{\alpha^2\nu}{M\omega_0^2},
\end{eqnarray}
with $\nu$ the density of states at the Fermi energy. {We will be particularly interested in the case in which $\lambda$
is  large compared to unity, so that although $T$ is small compared to  
$E_F$, $\lambda T$ 
can be larger
$E_F$ and even larger than the bandwidth $\Lambda$, {\it i.e.} there exists an interesting ``high temperature''  regime in which $E_F/\lambda \ll T \ll E_F$, where
  the quasiparticle scattering rate is larger than its energy, 
but the electrons  are nevertheless  quantum mechanically degenerate~\cite{Werman}.}

The current operator of this system is given by
\begin{eqnarray}
&&\mathbf{J}(i\omega_n)=e\sum_{a,\nu_n}\int \frac{d^dk}{(2\pi)^d} \mathbf{v}_{\mathbf{k}} c^\dagger_a(\mathbf{k},\nu_n)c_a(\mathbf{k},\nu_n+\omega_n)\notag\\
&&+\frac{e\alpha}{\sqrt{\beta N}}\sum_{a,b,\nu_n,\nu_m,r}\int \frac{d^dk d^dk'}{(2\pi)^{2d}} X^r_{ab}(\mathbf{k}-\mathbf{k'},\nu_n-\nu_m)\nonumber\\
&&\times \left(\frac{\partial g_r}{\partial \mathbf{k}}+ \frac{\partial g_r}{\partial \mathbf{k}'}\right)\left[c^\dagger_a (\mathbf{k},\nu_n)c_b(\mathbf{k'},\nu_m+\omega_n)+a\leftrightarrow b\right]\nonumber\\
&&\equiv \mathbf{J}^0 + \mathbf{J}^1;
\end{eqnarray}
here $\mathbf{v}_{\mathbf{k}}=\frac{\partial\epsilon}{\partial\mathbf{k}}$. {This can be derived, for instance, by coupling the electrons to a vector potential $\mathbf{A}$ by replacing $c_a(\mathbf{k}) \rightarrow c_a(\mathbf{k}-e\mathbf{A})$ in Eq.~(\ref{eq:model}), and differentiating the action with respect to $\mathbf{A}$.} $\mathbf{J}^0$ 
{derives directly from} the non-interacting electrons' kinetic energy, while $\mathbf{J}^1$ represents a phonon-assisted conductivity channel.

\emph{Single electron properties.--} Taking the limit $N\rightarrow \infty$ allows us to solve the model (\ref{eq:model}) order by order in $1/N$. Just as in Ref.~\cite{Raghu}, the full set of rainbow diagrams contributes to the electron self-energy to lowest order in $1/N$. This results
in a self-consistent Dyson's equation for the fermion self-energy:

\begin{eqnarray}
\label{eq:selfenergy}
\Sigma(\mathbf{k},\omega)=\frac{\lambda T}{\nu}\sum_r\int\frac{d^dk'}{(2\pi)^d}\frac{|g_r(\mathbf{k},\mathbf{k}')|^2}{\omega-\xi_{\mathbf{k'}}-\Sigma(\mathbf{k}',\omega)}.
\end{eqnarray}

For solids with a constant density of electrons, this equation must be solved simultaneously with the equation for the density per flavor
\begin{eqnarray}\label{eq:chempotential}
n =  \int \frac{d^d k}{(2\pi)^d} \langle  c^\dagger_\alpha (\mathbf{k}) c^{\vphantom{\dagger}}_\alpha (\mathbf{k}) \rangle 
\approx  \int \frac{d^d k}{(2\pi)^2} \int_{-\infty}^0\frac{d\omega}{2\pi} A (\mathbf{k},\omega),\notag\\
\end{eqnarray}
which fixes the temperature dependent chemical potential $\mu$. Here $A(k,\omega)=-2\mathrm{Im}\frac{1}{\omega-\xi_{\mathbf{k}}-\Sigma({ \mathbf{k},}\omega)}$ is the spectral function. For details of the solution, see Appendix; here we state the results.

At low temperatures, $\lambda T\ll E_F$, the chemical potential is approximately temperature-independent and the scattering rate on the Fermi surface, given by $1/\tau(\mathbf{k})=-\mathrm{Im}[\Sigma(\mathbf{k},\omega=0)]\equiv\Sigma''(\mathbf{k},\omega=0)$, rises linearly with temperature:  $1/\tau(\mathbf{k})=\pi \lambda T/\nu\sum_r \int_\mathbf{k'} |g_r(\mathbf{k},\mathbf{k}')|^2\delta(\xi_{\mathbf{k}'})$, with $\int_{\mathbf{k}}\equiv \int d^dk/(2\pi)^d$; this is the famous semiclassical result \cite{Ashcroft}. 

In the high temperature limit, $\lambda T\gg\Lambda$, the temperature dependence is given by (see details in the Appendix)
{
\begin{eqnarray}
\label{eq:selfenergyhighT}
&&\mu(T) = \tilde{\mu}_0\sqrt{\lambda T/\nu}\\
&&\Sigma(\mathbf{k},\omega;T) = \tilde{\Sigma}(\mathbf{k},\tilde{\omega})\sqrt{\lambda T/\nu}\notag
\end{eqnarray}
where $\tilde{\omega}=\omega/\sqrt{\lambda T/\nu}$, and $\tilde{\mu}_0, \tilde{\Sigma}(\mathbf{k},\omega)$ are found by solving the coupled, temperature independent equations
\begin{eqnarray}
\label{eq:dimensionless}
\tilde{\Sigma}(\mathbf{k},\tilde{\omega}) = \sum_r\int\frac{d^dk'}{(2\pi)^d}\frac{|g_r(\mathbf{k},\mathbf{k}')|^2}{\tilde{\omega}+\tilde{\mu}_0-\tilde{\Sigma}(\mathbf{k}',\tilde{\omega})},\\
n = \int\frac{d^dk'}{(2\pi)^d}\int _{-\infty}^{\tilde{\mu}_0}\frac{d\tilde{\omega}}{2\pi}\mathrm{Im}\left[\frac{1}{\tilde{\omega}-\tilde{\Sigma}(\mathbf{k}',\tilde{\omega})}\right].\notag
\end{eqnarray}
}
At high temperature, a crossover occurs to a square-root dependence of the self energy on the temperature. The crossover occurs around $\lambda T\approx\mu$.

\begin{figure}[h!]
\centering
\includegraphics[width=0.5\textwidth]{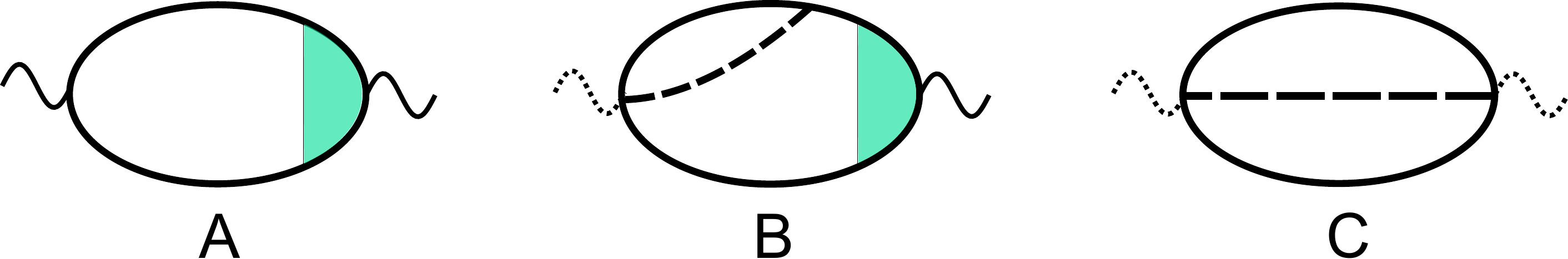}
  \caption{(Color online.) The three diagrams which contribute to the conductivity. Bold lines are renormalized electron propagators, dashed lines are phonon propagators, and the full and dashed wiggly lines correspond to $J^0$ and $J^1$, respectively. The colored area represents the renormalized $J^0$ vertex function.}
\label{fig:diagrams}
  \centering
    \end{figure}

\emph{Conductivity.--}
The {D.C.} conductivity is given by summing over three different channels (see Fig.~\ref{fig:diagrams}):
\begin{eqnarray}
&&\sigma=\sigma^{00}+2\sigma^{01}+\sigma^{11}\mbox{  , where}\\
&&\sigma^{ij}=-\lim_{\omega\to 0}\frac{\mathrm{Im} \Pi^{ij}(\omega)}{\omega},\notag \\
&&\Pi^{ij}(\omega)=\left\langle J_x^i(i\omega_n)J_x^j (-i\omega_n)\right\rangle|_{i\omega_n\rightarrow \omega+i\delta}\nonumber.
\end{eqnarray}

The full details of the calculation of the conductivity are given in the Appendix; here, for simplicity, we will sketch the calculations 
without vertex corrections. Vertex corrections are included in the figures, and do not change the behavior qualitatively.

$\sigma^{00}$ has been calculated in Ref.~\cite{Werman}; neglecting vertex corrections, it is given by
\begin{eqnarray}\label{eq:conductivity}
\sigma^{00}(T)&=&-\lim_{\omega\rightarrow 0}\frac{\mathrm{Im}\Pi^{00}(\omega_n\rightarrow \omega+i\delta,T)}{\omega}\nonumber\\
&=&-\lim_{\omega\rightarrow 0}\frac{e^2N}{\beta\omega}\mathrm{Im}\sum_{\nu_n}\int\frac{d^dk}{(2\pi)^d} v_\mathbf{k}^2 \nonumber\\
&&\times \mathcal{G}(i\nu_n,\mathbf{k})\mathcal{G}(i\nu_n+i\omega_n,\mathbf{k})|_{i\omega_n\rightarrow\omega+i\delta}\nonumber\\
&\approx&\frac{e^2N}{4\pi}\int\frac{d^dk}{(2\pi)^d}v_\mathbf{k}^2 \left[A(\mathbf{k},\omega=0)\right]^2.
\label{eq:conduct}
\end{eqnarray}
$\mathcal{G}(i\nu_n,k)$ is the fully dressed electron Green's function. In the last line of Eq.~(\ref{eq:conduct}), we have inserted the spectral representation of the Green's function, performed the Matsubara summation over $\nu_n$ (see, e.g.,~\cite{Mahan}), 
and used the fact that the Fermi function $n_F(\epsilon)$ obeys $\frac{dn_F(\epsilon)}{d\epsilon} \approx -\delta(\epsilon)$ in the regime $T\ll E_F$, assuming that $A(\mathbf{k},\omega)$ changes slowly on the scale of $T$~\footnote{This is justified because, at low temperature, $A(\mathbf{k},\omega)$ varies on the scale of $\Sigma''(\omega = 0,T \ll E_F/\lambda) \sim \lambda T$, assumed to be much larger than $T$. [Here we have assumed that the density of states, and hence $\Sigma''(\omega)$, vary slowly around zero energy on the scale of $T$.] At high temperature the spectral function varies on the scale of $\Sigma''(\omega=0,T \gg \Lambda /c) \sim \sqrt{\frac{\lambda T}{\nu}}\gg T$ [see Eq.~(\ref{eq:selfenergyhighT})].}. 

$\sigma^{11}$ is the channel responsible for resistivity saturation. To lowest order in $1/N$, the $\Pi^{11}$ correlation function is given by
\begin{eqnarray}
&&\Pi^{11}(i\omega_n)=e^2N{\alpha^2}\sum_r \int \frac{d^dkd^dk'}{(2\pi)^{2d}}  \left\lvert\frac{\partial g_r}{\partial \mathbf{k}}+ \frac{\partial g_r}{\partial \mathbf{k}'}\right\rvert^2 \nonumber \\
&&\times \frac{1}{\beta^2}\sum_{n,m}G(\mathbf{k},i\nu_n)G(\mathbf{k'},i\nu_m+i\omega_n)D(\mathbf{k}-\mathbf{k'},i\nu_n-i\nu_m), \nonumber \\
\end{eqnarray}
with $D(\mathbf{q}, i\omega_n)$ the phonon propagator, which is unrenormalized to lowest order in $1/N$. To leading order in $\omega_0/T$, this results in
\begin{eqnarray}
\label{eq:pi11}
&&\Pi^{11}(i\omega_n)=e^2N\frac{\lambda T}{\nu} \sum_r \int \frac{d^dkd^dk'}{(2\pi)^{2d}} \left\lvert\frac{\partial g_r}{\partial \mathbf{k}}+ \frac{\partial g_r}{\partial \mathbf{k}'}\right\rvert^2\nonumber \\ 
&&\times   \frac{1}{\beta}\sum_{n}G(\mathbf{k},i\nu_n)G(\mathbf{k'},i\nu_n+i\omega_n) \nonumber\\
&&=e^2N\frac{\lambda T}{\nu}\sum_r\int \frac{d^dkd^dk'}{(2\pi)^{2d}} \left\lvert\frac{\partial g_r}{\partial \mathbf{k}}+ \frac{\partial g_r}{\partial \mathbf{k}'}\right\rvert^2 \nonumber\\
&&\times \int \frac{d\epsilon_1}{2\pi} \frac{d\epsilon_2}{2\pi} A(\mathbf{k},\epsilon_1)A(\mathbf{k'},\epsilon_2)\frac{n_F(\epsilon_1)-n_F(\epsilon_2)}{i\omega_n+\epsilon_1-\epsilon_2} 
,
\end{eqnarray}
where we have 
 inserted the spectral representation and performed the Matsubara summation. Therefore, again using the fact that $T\ll E_F$,
\begin{eqnarray}
\label{eq:sigma11}
\sigma^{11}&=&\frac{e^2N\lambda T}{4\pi\nu }\sum_r \int \frac{d^dkd^dk'}{(2\pi)^{2d}} \left\lvert\frac{\partial g_r}{\partial \mathbf{k}}+ \frac{\partial g_r}{\partial \mathbf{k}'}\right\rvert^2 \nonumber\\
&\times& A(\mathbf{k},0)A(\mathbf{k'},0).
\end{eqnarray}

At high temperatures it is possible to approximate $A(\mathbf{k},0)={-2}\sqrt{\frac{\nu}{\lambda T}}\mathrm{Im}\frac{1}{\tilde{\mu}_0-\tilde{\Sigma}(\mathbf{k},0)}$, with both $\tilde{\Sigma}(\mathbf{k},0)$ and $\tilde{\mu}_0$ temperature independent. Therefore, at high $T$, $\sigma^{11}$ saturates to a temperature and coupling strength-independent value. 

A plot of $\sigma^{00}$ and $\sigma^{11}$, calculated for a two-dimensional tight binding model, is shown in Fig.~\ref{fig:conductivities}.  The contribution of the $\{01\}$ channel, $\sigma^{01}$ (calculated in the Appendix) is found to be negligible compared to $\mathrm{max}[\sigma^{00}, \sigma^{11}]$, both at low and high temperatures.

\begin{figure}	
\centering
\includegraphics[width=0.4\textwidth]{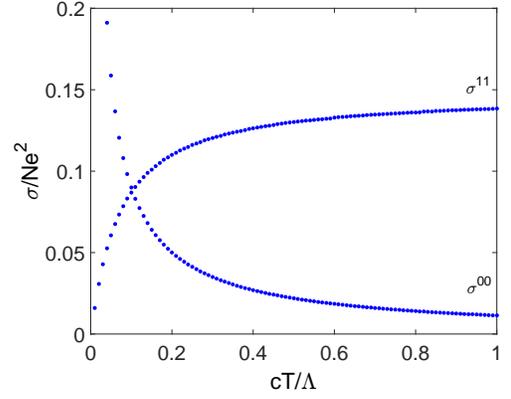}
  \caption{The $\sigma^{00}$ and $\sigma^{11}$ conductivities per band.  $\sigma^{00}$ 
falls as $1/T$ 
{(although with different proportionality constants) in both temperature ranges  $\omega_0 \ll T\ll\mu/\lambda$ 
and $\mu/\lambda\ll T \ll \mu$.  $\sigma^{11}$ grows
 in the lower range of $T$, then saturates as it approaches the value $e^2/2\pi$ in the higher.} This is calculated for a two dimensional tight binding model, in which the phonon displacement couples to the nearest-neighbor hopping amplitude of the electrons.}
\label{fig:conductivities}
  \centering 
\end{figure}
\emph{Resistivity.--} Adding the three conductivity channels, the resistivity $\rho = 1/\sigma$ of the model is shown in Fig.~\ref{fig:ressat}. 
For $\omega_0 \ll T\ll E_F/\lambda$, 
the $\{00\}$ channel dominates, and the resistivity rises linearly with temperature: 
\begin{eqnarray}
\rho(T \ll E_F/\lambda)  \approx 1/\sigma^{00} \approx \frac{2\pi}{e^2N}\frac{\lambda T}{\bar{v}_F^2\nu}
\end{eqnarray}
where $\bar{v}_F^2=1/\pi\int_{\mathbf{k}} \left[v_{\mathbf{k}}^2\delta(\xi_{\mathbf{k}}) /\sum_r \int_{\mathbf{k'}} |g_r(\mathbf{k},\mathbf{k'})|^2\delta(\xi_{\mathbf{k}'})\right]$.
This is the Bloch-Gr\"uneisen formula for $T>\omega_0$. 

\begin{figure}	
\centering
\includegraphics[width=0.4\textwidth]{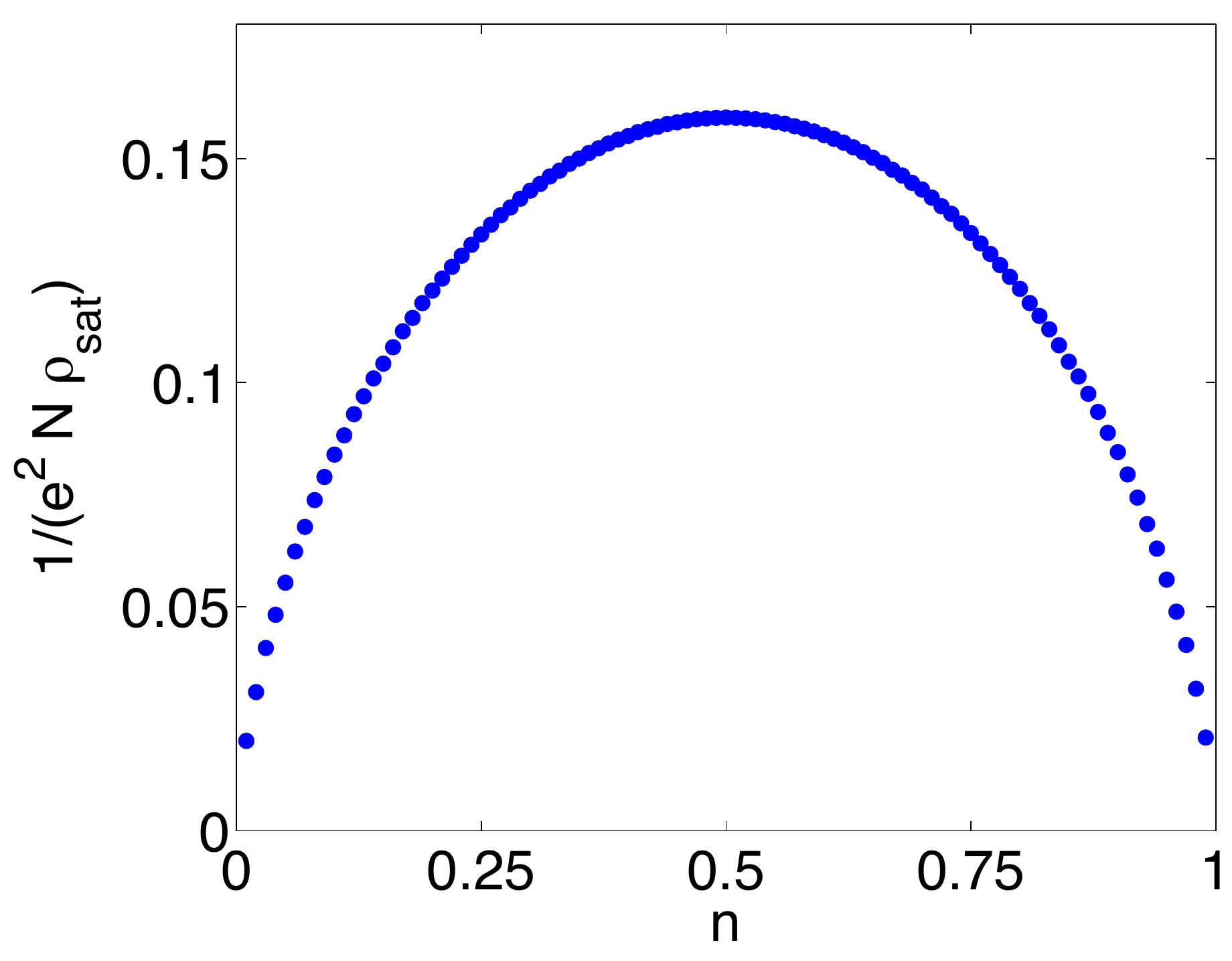}
  \caption{Saturation resistivity, $\rho_{\mathrm{sat}}$, as a function of the density of electrons per 
  {site per flavor.}}
\label{fig:rhosat}
  \centering 
\end{figure}

At high temperatures, however, the linear increase in $1/\sigma^{00}$ is offset by the parallel addition of the saturating $\{11\}$ channel, the rapid growth of the resistivity is checked, and the resistivity saturates at the value
\begin{eqnarray}
\rho_{\mathrm{sat}} &=&  \frac{\pi \hbar}{e^2N} \Bigg[ \sum_r\int\frac{d^dkd^dk'}{(2\pi)^{2d}} \left\lvert\frac{\partial g_r}{\partial \mathbf{k}} + \frac{\partial g_r}{\partial \mathbf{k}'}\right\rvert^2  \nonumber \\
&\times& \mathrm{Im} \frac{1}{\tilde{\mu}_0-\tilde{\Sigma}(\mathbf{k},0)}\mathrm{Im}\frac{1}{\tilde{\mu}_0-\tilde{\Sigma}(\mathbf{k}',0)} \Bigg]^{-1}
\label{eq:rhosat}
\end{eqnarray}
{Here, we have reintroduced $\hbar$ for clarity. From Eq.~(\ref{eq:rhosat}), it is clear that $\rho_{\mathrm{sat}}$ is independent of temperature and of the electron-phonon coupling strength, $\lambda$. It does, however, depend on the form factor $g_r(\mathbf{k}, \mathbf{k}')$ and on the electron density $n$ [through the dependence of $\tilde{\mu}$ and $\tilde{\Sigma}(\mathbf{k},0)$ on $n$, Eq.~(\ref{eq:dimensionless})]. In Fig.~\ref{fig:rhosat} we show $1/\rho_{\mathrm{sat}}$ as a function of $n$ in our model. $\rho_{\mathrm{sat}}$ reaches a minimum close to $h/(Ne^2)$ at half filling. Near $n=0$ and $n=1$, $\rho_{\mathrm{sat}}$ diverges~\footnote{Note that within our model, in order to access the saturation regime in the limit $n\rightarrow 0$ while keeping $T\ll E_F$, one has to take $\lambda \rightarrow \infty$ while keeping $\lambda E_F$ fixed.}.
}

 At low fillings, the rapid decrease of $\sigma^{00}$ causes the resistivity to overshoot the saturating value, and the parallel addition of the channels causes the resistivity to decrease with temperature. In that case, the high temperature saturation value is approached \emph{from above} (see Fig.~\ref{fig:ressat}). Such behavior has been observed in certain heavy fermion compounds~\cite{Buschow}.

\emph{Optical conductivity.--} The optical conductivity $\sigma(\omega)$ can give insights into the physics of saturating metals and of bad metals~\cite{Hussey, Gunnarsson}. In conventional metals, the optical conductivity displays a pronounced Drude peak at all accessible temperatures, while materials which approach the MIR limit have been argued to lose this coherent contribution. To gain further insights into the mechanism of the saturation in our model, we now examine the optical conductivity.

\begin{figure}	
\centering
\includegraphics[width=0.4\textwidth]{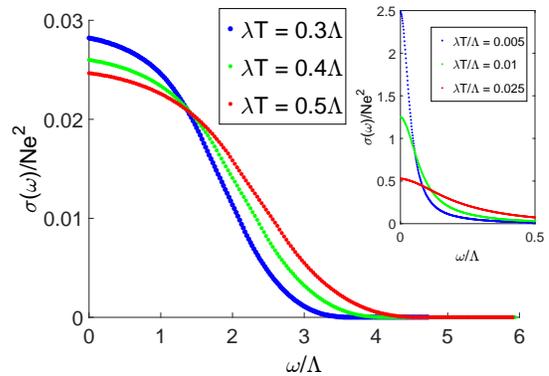}
  \caption{The optical conductivity $\sigma(\omega)$ for several temperatures. (inset) For low temperatures $\omega_0\ll T\ll E_F/\lambda$, a distinct Drude peak appears in the spectrum, of width $\lambda T$, and the conductivity vanishes for $\omega>\Lambda$. At high temperatures, the Drude peak is lost, but $\sigma(\omega)$ has a clear structure on the scale of $\sqrt{\lambda T/\nu}$; the optical conductivity has finite support over the effective bandwidth $\sqrt{\lambda T/\nu}\gg\Lambda$. In this model, resistivity saturation implies that the zeroth moment of the conductivity grows with temperature.}
\label{fig:optical}
  \centering 
\end{figure}

It is straightforward to extend the calculations described above to $\sigma(\omega)$ (see Appendix for details). The optical conductivity as a function of frequency for several temperatures is shown in Fig.~\ref{fig:optical}. At low temperatures, where $\sigma^{00}$ dominates, the conductivity shows a Drude peak whose width is proportional to $T$. 
At high temperatures (within the saturating regime), on the other hand, the optical conductivity consists of a broad peak whose height is nearly temperature independent, while \emph{its width increases with temperature}. This can be understood from the fact that, at asymptotically high temperatures, $\sigma(\omega)$ has support over a frequency range that scales with 
an effective bandwidth $\sqrt{\lambda T/\nu}$. 

Interestingly, this implies that the total spectral weight, defined as
\begin{eqnarray}
\mu_0 = \int_0^\infty \sigma(\omega)
\end{eqnarray}
increases with temperature. This is consistent with the sum rules concerning $\sigma(\omega)$, which within our model is given by
\begin{eqnarray}
\mu_0 = - \frac{\pi}{2}e^2 \big(\left\langle K_0 \right\rangle + \langle K_1 \rangle \big), 
\label{eq:sumrule}
\end{eqnarray}
where 
\begin{eqnarray}
K_0 &=& \sum_{\mathbf{k},a}  \frac{\partial^2 \xi_{\mathbf{k}}}{\partial k_x^2} c^\dagger_{a}(\mathbf{k}) c^{\vphantom{\dagger}}_{a}(\mathbf{k}), \\ 
K_1 &=& \frac{4\alpha}{\sqrt{N}} \sum_{\mathbf{k}, \mathbf{k}',a,b,r} \gamma_r(\mathbf{k}, \mathbf{k}') X^r_{a,b}(\mathbf{k} - \mathbf{k}' ) \nonumber \\ 
&\times& [c^\dagger_{a}(\mathbf{k}) c^{\vphantom{\dagger}}_{b}(\mathbf{k}) + h.c.]. \nonumber 
\end{eqnarray}
Here, $ \gamma_r(\mathbf{k},\mathbf{k'}) \equiv [\partial^2_{k_x} + \partial^2_{k_x'} + 2\partial_{k_x} \partial_{k_x'} ] g_r(\mathbf{k},\mathbf{k'})$. It is the second term in Eq.~(\ref{eq:sumrule}) that is responsible for the increase of the spectral weight, $\mu_0 \propto \sqrt{T}$, for $\lambda T \gg \Lambda$. 
This reflects the fact that in this regime, the phonon-assisted hopping channel dominates the transport. In real materials, however, this behavior 
{might be hard to observe, as it could } be masked by high-energy features in $\sigma(\omega)$ due to inter-band transitions. 

\emph{Numerics.--} The analytical results described above are confined to the $N\rightarrow\infty$ limit.  
One may then ask to what extent the physics of a system with a finite number of electronic bands and phonon modes is captured by the $N\rightarrow \infty$ picture. To assess this, we have performed numerical simulations of the model in Eq. (\ref{eq:model}) for finite values of $N$. The simulations are done by treating the phonons as a classical static field with a distribution that corresponds to the free energy of the system with a fixed phonon configuration, while the electrons are treated quantum mechanically. (See Appendix for additional details of the simulations.) The only approximation in this approach is to neglect the phonon dynamics, by taking $\omega_0 \rightarrow 0$. 
The problem can be solved fully quantum mechanically using quantum Monte Carlo (QMC), since the model~(\ref{eq:model}) does not suffer from a sign problem (although then, calculating the conductivity requires an analytic continuation to real time). Ref.~\cite{Calandra} demonstrated that at high temperatures, QMC results for a similar elecron-phonon model agree with the ``semiclassical'' approximation that neglects the phonon dynamics. 

In Fig.~\ref{fig:num}, the resistivity as a function of temperature is shown for systems with $N=2,4,6, 8$, along with the analytical $N \rightarrow \infty$ result. The numerical results approach the $N \rightarrow \infty$ curve, showing that the approach to the $N \rightarrow \infty$ limit is not singular. It is also clear that signatures of saturation appear already at small $N$, rendering our analysis pertinent for physical systems.
 
\emph{Discussion.--} 
It has been argued that saturation 
is connected with a limit quantum mechanics imposes on the maximal quasiparticle scattering rate, or equivalently on the minimal mean free path. In our model, the inverse electron lifetime increases without bound as $\Sigma''(\mathbf{k},0)\propto\sqrt{T}$; this is clearly \emph{not} the mechanism for saturation.  
{However, the origin of the saturation is quantum mechanical - it relies on the finite bandwidth $\Lambda$ of the system, and the saturation value is proportional to Planck's constant $h$ [Eq.~(\ref{eq:rhosat})]. }

{We can also address the question of whether the correct criterion for saturation is $\ell \approx a$ or $\ell \approx 2\pi/k_F$, by looking at the low density limit where $E_F \ll \Lambda$. In this regime, we find two distinct crossovers that occur when $T$ becomes comparable to $E_F/\lambda$ and $\Lambda/\lambda$, respectively. In the first of these crossovers, where the 
Boltzman approach breaks down, the slope of the linear increase of $\rho$ deviates from its low-$T$ value~\cite{Werman}; in the second crossover, where the extrapolated mean free path satisfies $\ell \approx a$, the saturation occurs. If $E_F$ and $\Lambda$ are parametrically different from each other (as in lightly doped semiconductors), 
 the resistivity may rise beyond the saturation value and then approach it from above (see Fig.~\ref{fig:ressat}).  
}
%
The value of the resistivity at saturation is $h a^{d-2}/e^2 $ times a numerical factor [
Eq.~(\ref{eq:rhosat})] that depends on the electron density and the 
 electron-phonon coupling form factor. The saturation value is 
 not universal, although it is independent of the overall electron-phonon coupling strength. 
%


\emph{Relation to other works.--} The $\sqrt{T}$-dependence of the self-energy at high temperatures has been found by Millis et al.~\cite{Millis} for an $N=1$ electron-phonon system, using DMFT. The importance of the coupling of the phonons to the electronic kinetic energy 
was recognized by Calandra et al.\cite{Calandra} 
They used quantum Monte Carlo (QMC) to compute the resistivity of a $5$-fold degenerate electron band coupled to optical phonons via the hopping matrix elements and 
observed resistivity saturation. 
  In contrast, 
  in a model in which the phonons couple to the site energies, 
the resistivity  did not saturate. 
  They also observed that the resistivity saturation depends on the number of degenerate electronic bands. 

Our analysis clearly elucidates why coupling to the kinteic energy is 
important; it is the conductance channel which originates form this coupling that causes the saturation. This gives a natural physical interpretation of the phenomenological parallel resistor formula. We note that the mechanism described in this work for resistivity saturation is different from the interpretation given in Ref.~\cite{Gunnarsson1}, which is based on the conductivity f-sum rule. In particular within our model, the integral of $\sigma(\omega)$ 
increases with temperature. {This is due to an increase of the effective bandwidth with temperature (see Fig.~\ref{fig:optical})}.   

\emph{Conclusions.--} We present a tractable electron-phonon model 
that displays resistivity saturation. At low temperatures, $\omega_0\ll T\ll E_F/\lambda$, the resistivity increases linearly with temperature, according to the semiclassical formula. At high temperatures, 
{ $T \gg \Lambda/\lambda$}, 
the resistivity saturates to a temperature and coupling strength-independent value. 
{The saturation is not a result of a limit on the scattering rate, but due to the existence of an additional phonon-assisted conductivity channel that becomes effective at higher temperature.  
This gives a natural microscopic interpretation for the phenomenological parallel resistor formula.}

Beyond the possible implications for the resistivity of metals, the analysis presented here, together with the one presented in Ref.~\cite{Werman}, provide examples of metallic transport in a regime that cannot be described in terms of coherent quasi-particles. It may be possible to extend this analysis, using an appropriate large-$N$ limit, to other problems of unconventional transport, e.g., where the scattering is dominated by electron-electron interactions. We leave such extensions to future work. 

\emph{Acknowledgements.--} We thank P. Allen, E. Altman, A. Auerbach, S. Hartnoll, and S. Raghu for illuminating discussions. E. B. and Y. W. were supported by the ISF under grant 1291/12, by the US-Israel BSF under grant 2014209, and by a Marie Curie CIG grant. S. K. was supported in part by NSF grant \#DMR 1265593 at Stanford.

\bibliography{paper}

\begin{widetext}
\appendix*
\section{Details of the calculation}
\subsection{Model} We use a two-dimensional tight binding model in which optical phonons are coupled to the hopping amplitude of the electrons:
\begin{eqnarray}
&&H=-t\sum_{a,i,r} \left(c^\dagger_{a,i} c_{a,i+r}+h.c.\right) \\
&&+ \sum_{a,b,i,r}\left[\frac{1}{2}M\omega_0^2\left(X^r_{ab,i}\right)^2+\frac{1}{2M}\left(P^r_{ab,i}\right)^2\right]\nonumber\\
&&+\frac{\alpha}{\sqrt{\beta N}}\sum_{a,b,i,r}X^r_{ab,i}\left(c^\dagger_{a,i} c_{b,i+r}+h.c.+a\leftrightarrow b\right).\nonumber
\end{eqnarray}
The corresponding Lagrangian is
\begin{eqnarray}
&&L=\sum_{a,\nu_n}\int \frac{d^2k}{(2\pi)^2} \left[i\nu_n-\xi_\mathbf{k}\right] c^\dagger_a(\mathbf{k},\nu_n)c_a(\mathbf{k},\nu_n)\\
&&+\sum_{a,b,\omega_n,r}\int \frac{ d^2q}{(2\pi)^2}\left[\frac{1}{2}M\omega_0^2+\frac{1}{2}M\omega_n^2\right]|X_{ab}^r(\mathbf{q},\omega_n)|^2\nonumber\\
&&+\frac{\alpha}{\sqrt{\beta N}}\sum_{a,b,\nu_n,\nu_m,r}\int \frac{ d^2k}{(2\pi)^2}\frac{ d^2k'}{(2\pi)^2} X_{ab}^r(\mathbf{k}-\mathbf{k'},\nu_n-\nu_m)\times\nonumber\\
&&\left[c^\dagger_a(\mathbf{k},\nu_n)c_b(\mathbf{k'},\nu_m)+a\leftrightarrow b\right]\times\left(e^{ik_ra}+e^{-ik'_ra}\right)\nonumber
\end{eqnarray}

where $c^\dagger_a(\mathbf{k},\nu_n)$ creates an electron of wavevector $\mathbf{k}$, Matsubara frequency $\nu_n$, and flavor $a$; $\xi_\mathbf{k}=\epsilon_\mathbf{k}-\mu$, with $\epsilon_\mathbf{k}=-2t\sum_r \cos(k_r )$, $r=\hat{x},\hat{y}$, and $\mu$ the chemical potential. $X_{ab}^r(\mathbf{q},\omega_n)$ is the fourier transform of the phonon displacement operator which lives on the $r$ links; the phonon has mass $M$ and optical frequency $\omega_0$. $\alpha$ is the electron-phonon coupling parameter. This model corresponds to $g_r(\mathbf{k},\mathbf{k}) = e^{ik_r}+e^{-ik'_r}$. In the following we take $M\rightarrow\infty$, which corresponds to the limit $\omega_0\rightarrow 0$ while keeping $K$, the phonon spring constant, finite.

The current operator is then given by
\begin{eqnarray}
&&J(i\omega_n)=\\
&&2ev_0\sum_{a,\nu_n}\int \frac{d^2k}{(2\pi)^2} \sin(k_x)c^\dagger_a(\mathbf{k},\nu_n)c_a(\mathbf{k},\nu_n+\omega_n)\nonumber\\
&&+\frac{e\alpha}{\sqrt{\beta N}}\frac{v_0}{t}\sum_{a,b,\nu_n,\nu_m}\int \frac{d^dk}{(2\pi)^d} \frac{d^dk'}{(2\pi)^d} X^r_{ab}(\mathbf{k}-\mathbf{k'},\nu_n-\nu_m)\nonumber\\
&&\left[c^\dagger_a(\mathbf{k},\nu_n)c_b(\mathbf{k'},\nu_m+\omega_n)+a\leftrightarrow b \right]\frac{1}{i}\left(e^{ik_r}-e^{-ik'_r}\right)\nonumber\\
&&\equiv J_r^0+J_r^1,\nonumber
\end{eqnarray}
with $v_0=\delta t$, $\delta$ being the lattice spacing. 

\subsection{Single electron properties} To leading order in $N^{-1}$, the self energy is given by the self consistency equation:
\begin{eqnarray}
\Sigma(\mathbf{k},\omega)=\frac{\alpha^2 T}{K}\sum_r\int \frac{d^2k'}{(2\pi)^2} \frac{[2+2\cos(k_r+k'_r)]}{\omega-\xi_{\mathbf{k}'}-\Sigma(\mathbf{k'},\omega)}
\end{eqnarray}
Since $\xi_\mathbf{k}$ and $\Sigma(\mathbf{k},\omega)$ are even functions of $\mathbf{k}$, it is possible to simplify
\begin{eqnarray}
&&\Sigma(\mathbf{k},\omega)=\frac{\alpha^2 T}{K}\sum_r\int \frac{d^2k'}{(2\pi)^2} \frac{[2+2\cos(k_r)\cos(k'_r)]}{\omega-\xi_{\mathbf{k}'}-\Sigma(\mathbf{k'},\omega)}\equiv \Sigma_I(\omega,T)+\Sigma_{II}(\omega,T)\sum_r\cos(k_r)\mbox{, with}\nonumber\\
&&\Sigma_I(\omega,T)=\frac{4\lambda T}{\nu}\int \frac{d^2k'}{(2\pi)^2} \frac{1}{\omega-\xi_{\mathbf{k}'}-\Sigma(\mathbf{k'},\omega)}=\nonumber\\
&&\frac{4\lambda T}{\nu}\int d^dk' \frac{1}{\omega-\xi_{\mathbf{k}'}-\Sigma_I(\omega,T)-\Sigma_{II}(\omega,T)\sum_r\cos(k'_r)}\nonumber\\
&&\mbox{and } \Sigma_{II}(\omega,T)=\frac{2\lambda T}{\nu}\int \frac{d^2k'}{(2\pi)^2} \frac{\cos(k'_x)}{\omega-\xi_{\mathbf{k}'}-\Sigma(\mathbf{k'},\omega)}=\nonumber\\
&&\frac{2\lambda T}{\nu}\int \frac{d^2k'}{(2\pi)^2}\frac{\cos(k'_x )}{\omega-\xi_{\mathbf{k}'}-\Sigma_I(\omega,T)-\Sigma_{II}(\omega,T)\sum_r\cos(k'_r)}.\nonumber\\
\end{eqnarray}

Equation~\ref{eq:chempotential} was solved by obtaining $\Sigma_{I}$ and $\Sigma_{II}$ iteratively for each $\omega$, and demanding the chemical potential $\mu$ satisfy
\begin{eqnarray}
&&n =  \int \frac{d^2 k}{(2\pi)^2} \int_{-\infty}^\mu \frac{d\omega}{2\pi}\mathrm{Im}\frac{1}{\omega+(2t-\Sigma_{II}(\omega))(\cos(k_x)+\cos(k_y))-\Sigma_I(\omega)}.
\end{eqnarray}

At high temperatures satisfying $\lambda T\gg\Lambda$, we can neglect $\epsilon_\mathbf{k}$ relative to the $T$-dependent chemical potential and to the self energy; this will be shown to be self consistent. In this case,
\begin{eqnarray}
&&\Sigma(\mathbf{k},\omega)=\frac{\alpha^2 T}{K}\sum_r\int \frac{d^2k'}{(2\pi)^2} \frac{[2+2\cos(k_r+k'_r)]}{\omega+\mu-\Sigma(\mathbf{k'},\omega)}\nonumber\\
&&n =  \int \frac{d^2 k}{(2\pi)^2} \int_{-\infty}^\mu \frac{d\omega}{2\pi} \mathrm{Im}\frac{1}{\omega-\Sigma(\mathbf{k},\omega)}.
\end{eqnarray}
We define the dimensionless quantities
\begin{eqnarray}
&&\tilde{\omega} = \omega/\sqrt{\lambda T/\nu}\\
&&\tilde{\Sigma}(\mathbf{k},\omega) =\Sigma(\mathbf{k},\omega)/ \sqrt{\lambda T/\nu}\nonumber\\
&&\tilde{\mu_0}=\mu(T) /\sqrt{\lambda T/\nu}.\nonumber
\end{eqnarray}
They satisfy the temperature independent equations in Eq.~\ref{eq:dimensionless}, and depend only on the density. $\tilde{\mu}_0$ and $\tilde{\Sigma}(\mathbf{k},\omega\ll\sqrt{\lambda T/\nu})$ are found to be numbers of order one, rendering our approxiamation consistent.

\subsection{$\sigma^{00}$} Next, we calculate the $J^{00}$ renormalized vertex. In the calculations of the conductivity, we calculate $\sigma_{xx}$, and therefore only consider $g_{r=x}(\mathbf{k},\mathbf{k}')$. In addition, we set the electron charge $e=1$ for simplicity. As shown in Fig.~\ref{fig:vertex}, only ladder diagrams contribute to the vertex to lowest order in $1/N$. 
\begin{figure}[h!]
\centering
\includegraphics[width=0.5\textwidth]{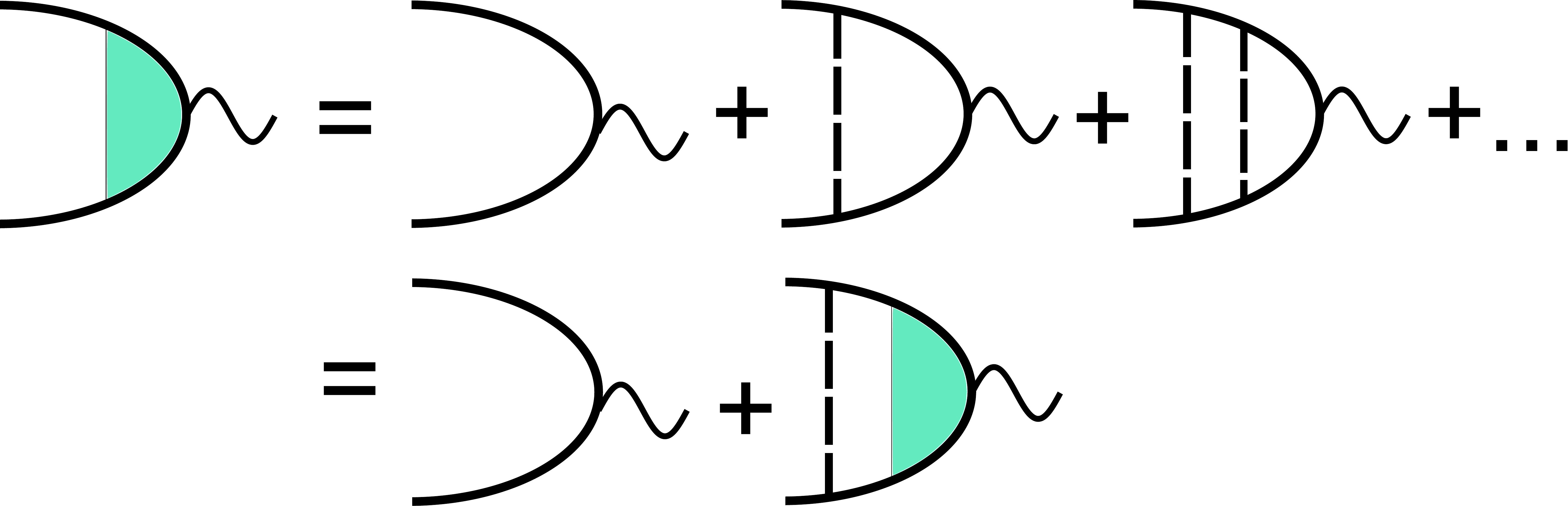}
  \caption{(Color online.) The $J^{00}$ vertex. Only ladder diagrams contribute to the vertex to lowest order in $1/N$.}
\label{fig:vertex}
  \centering
    \end{figure}
The ($x$-direction) vertex function $\Gamma(\mathbf{k},\nu_n,\nu_n+\omega_n)$ satisfies 
\begin{eqnarray}
&&\Pi^x(i\omega_n) = N\int\frac{d^2 k}{(2\pi)^2}\frac{1}{\beta}\sum_n G(\mathbf{k},\nu_n)G(\mathbf{k},\nu_n+\omega_n)\Gamma(\mathbf{k},\nu_n,\nu_n+\omega_n)\times 2t\sin(k_x a)\mbox{, with}\nonumber\\
&&\Gamma(\mathbf{k},\nu_n,\nu_n+\omega_n)=2t\sin(k_x a)+\nonumber\\
&&\frac{2\alpha^2}{K}\int\frac{d^2 k'}{(2\pi)^2}\frac{1}{\beta}\sum_m [1+\cos(k_x+k'_x)]D (\mathbf{k}-\mathbf{k'},\nu_n-\nu_m)G(\mathbf{k'},\nu_m)G(\mathbf{k'},\nu_m+\omega_n)\Gamma(\mathbf{k'},\nu_m,\nu_m+\omega_n)
\end{eqnarray}
Performing the Matsubara summation, to lowest order in $\omega_0/T$ only the pole at the phonon propagator is taken into account, and we are left with the self consistent equation
\begin{eqnarray}
&&\Gamma(\mathbf{k},\nu_n,\nu_n+\omega_n)=2t\sin(k_x a)+\\
&&\frac{2\lambda T}{\nu}\int\frac{d^2 k'}{(2\pi)^2} [1+\cos(k_x+k'_x)]G(\mathbf{k'},\nu_n)G(\mathbf{k'},\nu_n+\omega_n)\Gamma(\mathbf{k'},\nu_n,\nu_n+\omega_n)\nonumber
\end{eqnarray}

Using the ansatz $\Gamma(\mathbf{k},\nu_m,\nu_m+\omega_n) = \gamma(\nu_n,\nu_n+\omega_n)\sin(k_x)$, we get (using the fact that $G(\mathbf{k},\nu_n)$ is an even function of $k_x$)
\begin{eqnarray}
\label{eq:vertex}
\gamma(\nu_n,\nu_n+\omega_n) = 2t-\frac{2\lambda T}{\nu}\gamma(\nu_n,\nu_n+\omega_n)
\int\frac{d^2 k}{(2\pi)^2}G(\mathbf{k},\nu_n)G(\mathbf{k},\nu_n+\omega_n)\sin^2(k_x)\nonumber\\
= \frac{2t}{1+\frac{2\lambda T}{\nu}F(\nu_n,\nu_n+\omega_n)},
\end{eqnarray}
with
\begin{eqnarray}
F(\nu_n,\nu_n+\omega_n)=\int\frac{d^2 k}{(2\pi)^2}G(\mathbf{k},\nu_n)G(\mathbf{k},\nu_n+\omega_n)\sin^2(k_x)\nonumber\\
\end{eqnarray}

We now calculate $\Pi^{00}(i\omega_n)$:
\begin{eqnarray}
&&\Pi^{00}(i\omega_n) = N\int\frac{d^2 k}{(2\pi)^2}\frac{1}{\beta}\sum_n G(\mathbf{k},\nu_n)G(\mathbf{k},\nu_n+\omega_n)\times
2t\sin(k_x)\Gamma(\mathbf{k},\nu_n,\nu_n+\omega_n)\nonumber\\
&&=N(2t)^2\frac{1}{\beta}\sum_n\frac{F(\nu_n,\nu_n+\omega_n)}{1+\frac{2\lambda T}{\nu}F(\nu_n,\nu_n+\omega_n)}\equiv N\frac{(2t)^2}{\beta}\sum_n P(\nu_n,\nu_n+\omega_n)\nonumber
\end{eqnarray}

We perform the Matsubara summation following~\cite{Mahan}, using the usual contour integral method in the complex plane. The complex function $G(\mathbf{k},z)$, which satisfies ($\delta$ is an infinitesimal)
\begin{eqnarray}
&&G(\mathbf{k},z=\omega+i\delta) = G^R(\mathbf{k},\omega)\\
&&G(\mathbf{k},z=\omega-i\delta) = G^A(\mathbf{k},\omega) =G^R(\mathbf{k},\omega)^* \nonumber
\end{eqnarray}
 has a branch cut on the real axis. Therefore, $P(z,z+\omega_n)$ has branch cuts at $\Re[z]=0,-\omega_n$. Performing this integral (and using the fact that $n_F(\epsilon-i\omega_n) = n_F(\epsilon))$ results in
\begin{eqnarray}
\label{eq:summation}
&&\frac{\Pi^{00}(i\omega_n)}{(2t)^2N}=\frac{1}{\beta}\sum_nP(i\nu_n,i\nu_n+i\omega_n)=\\
&&-\int \frac{d\epsilon}{2\pi i} n_F(\epsilon)\left[ P(\epsilon+i\delta,\epsilon+i\omega_n)-P(\epsilon-i\delta,\epsilon+i\omega_n)\right]-\int \frac{d\epsilon}{2\pi i}n_F(\epsilon)\left[P(\epsilon-i\omega,\epsilon+i\delta)-P(\epsilon-i\omega,\epsilon-i\delta)\right]\nonumber
\end{eqnarray}

Therefore, using the fact that $T\ll\mu$, and thus $dn_F(\epsilon)/d\epsilon \approx -\delta(\epsilon)$,
\begin{eqnarray}
\label{continuation}
\sigma^{00} = -\frac{1}{\omega}\lim_{\omega\to 0}\mathrm{Im}\Pi^{00}(i\omega_n\rightarrow\omega+i\delta)
=N\frac{(2t)^2}{2\pi}\left[P(-i\delta,i\delta)-\Re\left[P(i\delta,i\delta)\right]\right],
\end{eqnarray}
where
\begin{eqnarray}
\label{eq:Pfun}
&&P(-i\delta,i\delta) = \frac{F(-i\delta,i\delta)}{1+\frac{2\lambda T}{\nu}F(-i\delta,i\delta)},\\
&&P(i\delta,i\delta) = \frac{F(i\delta,i\delta)}{1+\frac{2\lambda T}{\nu}F(i\delta,i\delta)},\nonumber\\
&&F(-i\delta,i\delta) = \int\frac{d^2k}{(2\pi)^2}G^A(\mathbf{k},0)G^R(\mathbf{k},0)\sin^2(k_x),\nonumber\\
&&F(i\delta,i\delta) = \int\frac{d^2k}{(2\pi)^2}G^R(\mathbf{k},0)G^R(\mathbf{k},0)\sin^2(k_x);\nonumber
\end{eqnarray}
these are easily computed once we have $\Sigma_I(0)$ and $\Sigma_{II}(0)$.

\subsection{$\sigma^{01}$} The ${01}$ current-current correlation function is given, to lowest order in $1/N$, by
\begin{eqnarray}
&&\Pi^{01}(i\omega_n)=-i\alpha^2N\frac{1}{\beta^2}\sum_{n,m}\int \frac{d^2k}{(2\pi)^2}\frac{d^2k'}{(2\pi)^2}D(\mathbf{k}-\mathbf{k'})\times\nonumber\\
&&G(\mathbf{k'},\nu_m)G(\mathbf{k},\nu_n)G(\mathbf{k},\nu_n+\omega_n)\Gamma(\mathbf{k},\nu_n,\nu_n+\omega_n)\left(e^{ik_x}-e^{-ik'_x}\right)\left(e^{-ik_x}+e^{ik'_x}\right).\nonumber\\
\end{eqnarray}
We perform the summation over $\nu_m$; to lowest order in $\omega_0/T$, this gives us (using the fact the all Green's functions are even in $k_x$, and inserting the vertex function~\ref{eq:vertex}):
\begin{eqnarray}
&&\Pi^{01}(i\omega_n)=N(2t)^2\frac{\lambda T}{\nu t}\frac{1}{\beta}\sum_{n}\int \frac{d^2k}{(2\pi)^2}\frac{d^2k'}{(2\pi)^2}\sin^2(k_x)\times\nonumber\\
&&G(\mathbf{k'},\nu_n)G(\mathbf{k},\nu_n)G(\mathbf{k},\nu_n+\omega_n)
\frac{1}{1+\frac{2\lambda T}{\nu}F(\nu_n,\nu_n+\omega_n)}\equiv N(2t)^2 \frac{\lambda T}{\nu t}\frac{1}{\beta}\sum_{n}R(\nu_n,\nu_n+\omega_n).
\end{eqnarray}
Note that to this order, the $J_1$ vertex is not renormalized.

Just as in the discussion above Eq.~\ref{eq:summation}, the complex function $R(z,z+\omega_n)$ has branch cuts at $\Re[z]=0,-\omega_n$; performing the Matsubara summation as in Eq.~\ref{eq:summation}, we get
\begin{eqnarray}\label{eq:summation2}
&&\Pi^{01}(i\omega_n)/\left[N(2t)^2\frac{\lambda T}{4\nu t}\right]=\frac{1}{\beta}\sum_nR(i\nu_n,i\nu_n+i\omega_n)=\\
&&-\int \frac{d\epsilon}{2\pi i} n_F(\epsilon)\left[ R(\epsilon+i\delta,\epsilon+i\omega_n)-R(\epsilon-i\delta,\epsilon+i\omega_n)\right]-\int \frac{d\epsilon}{2\pi i}n_F(\epsilon)\left[R(\epsilon-i\omega,\epsilon+i\delta)-R(\epsilon-i\omega,\epsilon-i\delta)\right]\nonumber
\end{eqnarray}
Therefore, in a procedure similar to that of Eq.~\ref{eq:summation}, we get
\begin{eqnarray}
\sigma^{01}=N\frac{\lambda T}{\nu t}\frac{(2t)^2}{2\pi}\Re\left[R(-i\delta,i\delta)-R(i\delta,i\delta)\right],
\end{eqnarray}
with
\begin{eqnarray}
&&R(-i\delta,i\delta) =\int \frac{d^2k}{(2\pi)^2}\frac{d^2k'}{(2\pi)^2}\sin^2(k_x)
\frac{G^A(\mathbf{k'},0)G^A(\mathbf{k},0)G^R(\mathbf{k},0)}{1+\frac{\lambda T}{2\nu}F(-i\delta,i\delta)}=P(-i\delta,i\delta)\int \frac{d^2k'}{(2\pi)^2}G^A(\mathbf{k'},0)\nonumber\\
&&R(i\delta,i\delta) = \int \frac{d^2k}{(2\pi)^2}\frac{d^2k'}{(2\pi)^2}\sin^2(k_x)
\frac{G^R(\mathbf{k'},0)G^R(\mathbf{k},0)G^R(\mathbf{k},0)}{1+\frac{\lambda T}{2\nu}F(i\delta,i\delta)}=P(i\delta,i\delta)\int \frac{d^2k'}{(2\pi)^2}G^R(\mathbf{k'},0),\nonumber\\
\end{eqnarray}
where the $P$ functions are defined in Eq.~\ref{eq:Pfun}.
Again, these can be computed easily once the self energy is obtained.
\subsection{$\sigma^{11}$}Finally, we turn to compute the conductivity in the ${11}$ channel. The $J_1$ vertex is not renormalized to lowest order in $1/N$, and Eq~\ref{eq:pi11} and~\ref{eq:sigma11} are exact to this order. We therefore substitute
\begin{eqnarray}
g_x(\mathbf{k},\mathbf{k'})=e^{ik_x}+e^{-ik'_x}
\end{eqnarray}
in these equations to get
\begin{eqnarray}
&&\sigma^{11}=N\frac{\lambda T}{2\pi\nu}\int \frac{d^2k}{(2\pi)^2}\frac{d^2k'}{(2\pi)^2}[1-\cos(k_x)\cos(k'_x)]A(\mathbf{k},0)A(\mathbf{k'},0).
\end{eqnarray}

\subsection{Optical conductivity}
Using eqns.\ref{eq:summation} and \ref{eq:summation2}, we get
\begin{eqnarray}
\sigma^{00}(\omega) = -\frac{N}{\omega}\mathrm{Im}\Pi^{00}(i\omega_n\rightarrow \omega+i\delta)=N\frac{(2t)^2}{2\pi}\Re\int_{-\omega}^0 d\epsilon \left[P(\epsilon-i\delta,\epsilon+\omega+i\delta) -P(\epsilon+i\delta,\epsilon+\omega+i\delta)\right],\nonumber\\
\sigma^{01} (\omega)= -\frac{N}{\omega}\mathrm{Im}\Pi^{01}(i\omega_n\rightarrow \omega+i\delta)=N\frac{\lambda T}{\nu t}\frac{(2t)^2}{2\pi}\Re\int_{-\omega}^0 d\epsilon \left[R(\epsilon-i\delta,\epsilon+\omega+i\delta) -R(\epsilon+i\delta,\epsilon+\omega+i\delta)\right],
\end{eqnarray}
while 
\begin{eqnarray}
&&\sigma^{11}(\omega)=\frac{\lambda Ta^2}{2\pi\nu}\int \frac{d^2k}{(2\pi)^2}\frac{d^2k'}{(2\pi)^2}[1+\cos(k_x)\cos(k'_x)]\int_{-\omega}^0 d\epsilon
A(\mathbf{k},\epsilon)A(\mathbf{k'},\epsilon+\omega);
\end{eqnarray}
where we have again used the fact that $T\ll E_F$.

\subsection{Numerical simulation}
In the numerical simulation, we consider a two dimensional tight binding model of size $L\times L$, and $N$ flavors of electrons. The single-particle Hamiltonian is given by
\begin{eqnarray}
&&H = H_0+H_{el-ph}\\
&&H_0 = -t\sum_{i,j,\alpha,\eta=\pm1}\left(|i,j,\alpha\right\rangle\left\langle i+\eta,j,\alpha|+|i,j,\alpha\right\rangle\left\langle i,j+\eta,\alpha|\right)
-t'\sum_{i,j,\alpha,\eta,\eta'=\pm1}|i,j,\alpha\rangle\langle i+\eta,j+\eta',\alpha|\nonumber\\
&&H_{el-ph} = -\frac{\alpha}{\sqrt{N}}\sum_{i,j,\alpha}\left[X^1_{i,j,\alpha,\beta}|i,j,\alpha\right\rangle\left\langle i+1,j,\beta|+X^2_{i,j,\alpha,\beta}|i,j,\alpha\right\rangle\left\langle i,j+1,\beta|\right]+h.c.\nonumber
\end{eqnarray}
where $|i,j,\alpha\rangle$ is the state with an electron of flavor $\alpha\in[1,N]$ at site ${i,j}$. $t$ is the nearest neighbor hopping parameter, while $t'$ represents next nearest neighbor hopping. $\alpha$ is the electron-phonon coupling strength, and $X^r_{\alpha,\beta,i,j}$ are the phonon modes at each site; they are characterized by the spring constant $K$. We choose $t'=-t$ and scale all energies by $t$. In addition, we work at filling $n=0.4$; these were chosen in order to minimize the susceptibility to lattice instabilities. Periodic boundary conditions were imposed.

The phonons are treated as classical fields and have no dynamics. This is justified in the limit $T\gg\omega_0$, where the quantum mechanical nature of the phonons is insignificant. However, we do consider the backaction of the electrons on the phonons. In our Monte Carlo simulation, at each step a single 
$X^r_{\alpha,\beta,i,j}$ is changed, and the resulting free energy is computed by
\begin{eqnarray}\label{eq:freeenergy}
F = \frac{1}{2}K\sum_{i,j,\alpha,\beta,r}\left(X^r_{i,j,\alpha,\beta}\right)^2-T\sum_n \log\left(1+e^{(-\epsilon_n-\mu)/T}\right),
\end{eqnarray}
where $\epsilon_n$ are the $L\times L\times N$ eigenvalues of the single-particle Hamiltonian for the given phonon configuration, and $\mu$ is the chemical potential obtained by demanding constant filling. The $\epsilon_n$'s are obtained by exact diagonalization.
The fact that the electrons modify the phonon configuration via the second term in \ref{eq:freeenergy}, allows us to observe lattice (Pierels) instabilities which occurs at strong couplings and low temperatures.

Each frozen phonon configuration represents a free electron system. It is therefore possible to calculate the conductivity by defining the ($x$-direction) single particle current operator
\begin{eqnarray}
J_x = -i\left[-t\sum_{i,j,\alpha,\eta=\pm1}\eta|i,j,\alpha\rangle\langle i+\eta,j,\alpha|
-\frac{\alpha}{\sqrt{N}}\sum_{i,j,\alpha,\beta,\eta=\pm1}X^1_{i,j,\alpha,\beta}\eta|i,j,\alpha\rangle\langle i+\eta,j,\beta|\right],
\end{eqnarray}
and calculating the optical conductivity as
\begin{eqnarray}
\sigma(\omega)=\frac{\pi}{L^2\omega}\sum_{n,n'}|\langle n |J_x|n'\rangle|^2\left[n_F(\epsilon_n)-n_F(\epsilon_{n'})\right]\delta(\omega+\epsilon_n-\epsilon_{n'}),
\end{eqnarray}
with $|n\rangle$ the eigenstate corresponding to $\epsilon_n$; again, both are obtained by exact diagonalization. We broaden the $\delta-$ function by using
\begin{eqnarray}
\delta(\epsilon)\approx-\pi\mathrm{Im}\frac{1}{\epsilon+0.01i}
\end{eqnarray}

In order to reduce finite size effects, we insert a flux quantum of $2\pi$ through the system. The effect of this flux is to break the point group and translation symmetries of the problem, which facilitates the convergence to the thermodynamic results~\cite{assaad2002depleted}. A gauge choice that describes a single flux quantum through a square lattice with a first and second neighbor hopping appears in Ref.~\cite{Berg}. For $N=8$ we use an $8\times 8$ lattice, and a finite size gap appears in the form of a dip in $\sigma(\omega\rightarrow 0)$, as shown in Fig.~\ref{fig:finitesize}. We ascertain that this dip vanishes as $L$ is increased, and is not due to an instability, such as a Pierels gap. We therefore approximate the conductivity as $\sigma_{dc} \approx \sigma(\omega = 0.0083
\Lambda)$.

\begin{figure}
\centering
\includegraphics[width=0.9\textwidth]{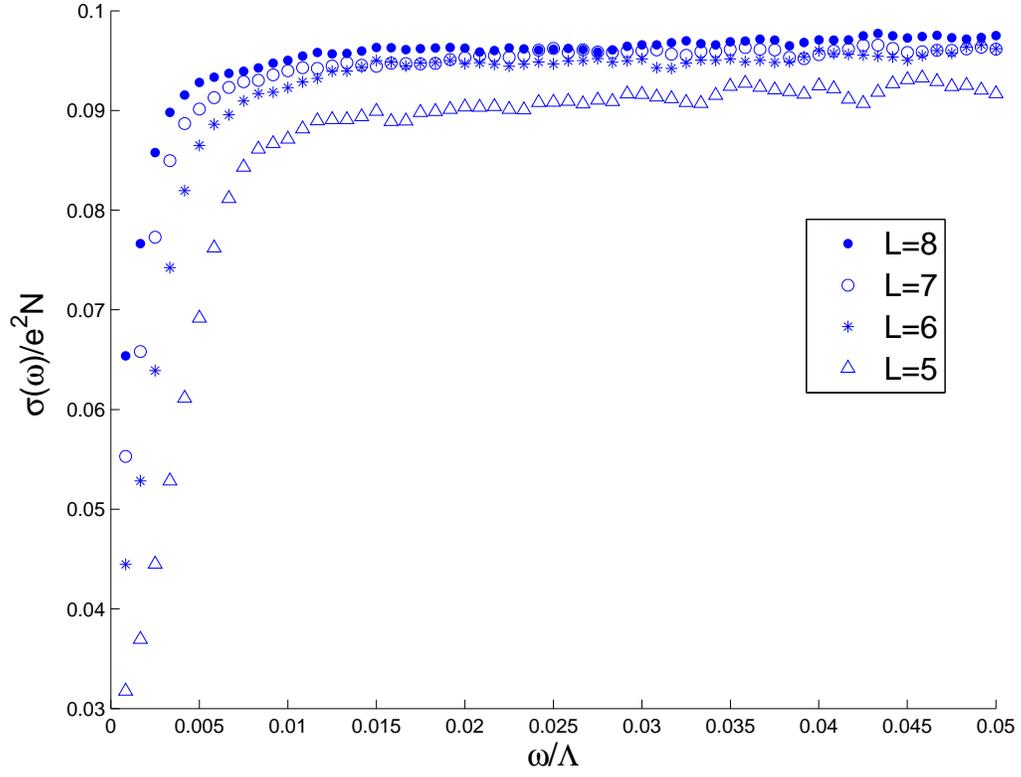}
  \caption{(Color online.) The optical conductivity for $N=8$ and several system sizes. The dip in the optical conductivity at low frequencies is due to a finite size gap, as shown by its decrease with increasing system size.}
\label{fig:finitesize}
  \centering
    \end{figure}

The system sizes we use are $L=8$ for $N=8$, $L=10$ for $N=6$, $L=12$ for $N=4$, and $L=17$ for $N=2$. In each simulation, 80 sweeps of the entire system were performed for thermalization, and 900 calculations of $\sigma(\omega)$ were averaged. We choose the coupling $\alpha$ such that
the dimensionless parameter
\begin{eqnarray}
\tilde{c}\equiv\frac{\alpha^2}{Kt}=3,
\end{eqnarray}
which is large enough to observe saturation at $T\ll E_F$, while small enough so that there is no Pierels transition at the temperatures we consider.
\end{widetext}

\end{document}